\documentclass{article}
\usepackage{graphicx,hyperref}
\pdfoutput=1

\title{Predictively Consistent Prior Effective Sample Sizes}

 \author{Beat
   Neuenschwander, 
Sebastian Weber, 
Heinz Schmidli,  Anthony O'Hagan
}

\date{July 12, 2019}

\begin{document}

\maketitle

\begin{abstract}
  Determining the sample size of an experiment can be challenging,
  even more so when incorporating external information via a prior
  distribution. Such information is increasingly used to reduce the
  size of the control group in randomized clinical trials. Knowing the
  amount of prior information, expressed as an equivalent prior
  \emph{effective sample size (ESS)}, clearly facilitates trial
  designs. Various methods to obtain a prior's \emph{ESS} have been
  proposed recently. They have been justified by the fact that they give
  the standard \emph{ESS} for one-parameter exponential
  families. However, despite being based on similar information-based
  metrics, they may lead to surprisingly different \emph{ESS} for
  non-conjugate settings, which complicates many designs with prior
  information. We show that current methods fail a basic
  predictive consistency criterion, which requires the expected
  posterior--predictive \emph{ESS} for a sample of size $N$
  to be the sum of the prior \emph{ESS} and $N$.  The \emph{expected
    local-information-ratio} \emph{ESS} is introduced
  and shown to be predictively consistent. It corrects the
  \emph{ESS} of current methods, as shown for normally distributed
  data with a heavy-tailed Student-t prior and exponential data with a
  generalized Gamma prior. Finally, two applications are discussed:
  the prior \emph{ESS} for the control group derived from
  historical data, and the posterior \emph{ESS} for
  hierarchical subgroup analyses.
\end{abstract}

Keywords: co-data, Fisher information, historical data, meta-analytic-predictive prior
distribution, prior predictive distribution

\maketitle

\section{Introduction}
\label{s:intro}
Sample sizes are an integral part of clinical trial designs and
usually follow from error rate (type-I, power) or precision
requirements. Such sample size determinations are standard if no
trial-external information is formally included in the analysis of the
parameter of interest.

If trial-external information contributes to the inference, one would
ideally want to quantify it via an equivalent \emph{effective sample
  size (ESS)}. Yet this can be difficult. For example, if historical
control data inform the prior distribution for the response rate of
the control group in a randomized trial, the amount of prior
information is not simply the number of historical control
subjects. It must be less due to between-trial heterogeneity, which is
unknown.

In health-care applications, additional data (or co-data, 
Neuenschwander, Roychoudhury, Schmidli (2016))
  are increasingly valued. In addition to the above example,
  applications include medical device trials (FDA (2010)),
  non-inferiority trials with historical or even
  concurrent placebo (FDA (2010)), pediatric trials with
  adult data (FDA (2004), Goodman and Sladky (2005)), health-technology assessments (Dias, Welton,
  Sutton, Ades (2011), Spiegelhalter, Abrams, Myles (2004)), 
  pharmacometrics (Demin, Hamren, Luttringer, Pillai et al. (2012), 
  Nedelman, Bretz, Fisch, Georgieva et al.  (2010)), and
  bridging studies.

In Section \ref{sec::methodology} we will review the standard
\emph{ESS} for the one-parameter exponential family, discuss current
methods for non-conjugate settings (Malec (2001), Morita,
Thall, M\"{u}ller (2008), Neuenschwander, Capkun-Niggli, Spiegelhalter (2010),
Pennello and Thompson (2008)), introduce the \emph{expected local-information ratio
  $ESS_{ELIR}$} as an alternative, investigate the different
\emph{ESS} for two examples, and show that only $ESS_{ELIR}$ is
predictively consistent. In Section \ref{sec::applications}, prior and
posterior  $ESS_{ELIR}$ will be discussed in the context of two recent
phase II trials.

\section{Methodology}
\label{sec::methodology}
In this Section, we aim to quantify the information for the parameter
$\theta$ of a statistical model $f(Y \vert \theta)$, expressed as an
equivalent \emph{effective sample size (ESS)}. The information about
$\theta$ is given probabilistically as a prior (or posterior)
distribution $p(\theta)$. The discussion will be restricted to
one-dimensional parameters.

\subsection{Effective sample sizes under conjugacy}
\label{sec::conjugacy}
Prior effective sample sizes are well understood for conjugate
one-parameter exponential families, such as: normal data with mean
$\mu$ (and known variance $s^2$) and a normal prior with variance
$s^2/n_0$ ($ESS=n_0$); binary data with response probability $\theta$
and a $\mbox{Beta}(a,b)$ prior ($ESS=a+b$); and, Poisson data with
mean (hazard) $\theta$ and a $\mbox{Gamma}(a,b)$ prior ($ESS=b$).

These \emph{ESS} can be motivated in various ways.  First, in the
updating rule from prior to posterior parameters, the sample size $n$
appears explicitly, suggesting a corresponding prior \emph{ESS}. For
example, for Poisson data with a $\mbox{Gamma}(a,b)$ prior, the second
parameter of the posterior Gamma distribution is $b+n$, implying $b$
as the prior \emph{ESS}.

Second, the posterior mean is a weighted average of the prior mean and
the standard parameter estimate, with weights proportional to the
prior \emph{ESS} and the sample size $n$. Again, for Poisson data, the
prior mean and parameter estimate are $a/b$ and $\sum Y_j/n$, and the
posterior mean $(a+\sum Y_j)/(b+n)$ is the weighted average of the
two, with weights proportional to $b$ and $n$. Of note, for
exponential data with mean $\mu$ and an $\mbox{inverse-Gamma}(a,b)$
prior, however, the \emph{ESS} for the weighted-mean approach is $a-1$
$(a>1)$, slightly different from the above updating-rule
\emph{ESS=a}. Moreover, for exponential data with censoring,
\emph{ESS} refers to the effective number of events rather than number
of observations.

\subsection{Variance- and precision-ratio methods}
A third, information-based justification is less well-known but will
serve as a basis for approaches to \emph{ESS} beyond conjugacy. It
relates the variance (or precision) of the prior to the variance of
an estimator $Y_N$ for $\theta$ from a sample of size $N$. The 
\emph{ESS} is then the $N$ for which the two variances are the same. Since
the variance of $Y_N$ will usually depend on $\theta$, the
expected variance under $p(\theta)$ is taken instead, which leads to
\begin{equation}
\label{ess:varratio1}
ESS = \frac{N \, E_{\theta} \lbrace \mbox{Var}(Y_N\vert\theta)\rbrace}{\mbox{Var}(\theta)}
\end{equation}
It can be shown that (\ref{ess:varratio1}) gives the standard
\emph{ESS} for the main one-parameter exponential families.
In the sequel, we will use a small modification, which will be needed
for the \emph{ESS} of Sections \ref{sec::mtm} and \ref{sec::elir}.
Letting $i_F(\theta)$ and $i_F(Y_1;\theta)$ be the expected and
observed Fisher information for one information unit,
\[
i_F(\theta) = E_{Y_1 \vert \theta} \left\{ i_F(Y_1;\theta) \right\}  =
-E_{Y_1 \vert \theta}
\left\{
\frac{d^2 \log f(Y_1 \vert \theta)}{d\theta^2} 
\right\}
\]
the variance-ratio and precision-ratio \emph{ESS}
are defined as
\begin{equation}
\label{ess:vr::pr}
ESS_{VR} = \frac{E_{\theta}\lbrace i_F^{-1}(\theta)\rbrace}{\mbox{Var}(\theta)}, \quad
ESS_{PR} = \frac{\mbox{Var}^{-1}(\theta)}{E_{\theta}\lbrace
  i_F(\theta) \rbrace}
\end{equation}
$ESS_{VR}$ and $ESS_{PR}$ are equal or close to the standard
\emph{ESS} for the main one-parameter exponential families. For
example, for Poisson data and a $\mbox{Gamma}(a,b)$ prior,
$\mbox{Var}(\theta)=a/b^2$, $i_F(\theta) = 1/\theta$, $E \lbrace
i_F^{-1}(\theta) \rbrace =a/b$, and $ESS_{VR}=b$. On the other hand,
$E \lbrace i_F(\theta)\rbrace =b/(a-1)$, and $ESS_{PR}=b (a-1)/a$,
which will be close to $b$ except for very small $a$.

In the sequel, we will use $ESS_{VR}$ and $ESS_{PR}$ in
(\ref{ess:vr::pr}) to represent the variance-ratio and precison-ratio
methods.  However, other variance-ratio methods have been suggested by
Malec (2001), Neuenschwander et al. (2010), and Pennello and Thompson
(2008). They obtain the \emph{ESS} of a prior by relating its
variance to the variance from an analysis for which the \emph{ESS} is
known. We do not include these proposals in the following comparisons
because they are similar to $ESS_{VR}$ and $ESS_{PR}$.

\subsection{The Morita-Thall-M\"{u}ller (MTM) method}
\label{sec::mtm}
Another, more involved information-based \emph{ESS} has been suggested
in the seminal paper by Morita et al.  (2008). In addition
to the Fisher information, it uses the information of the prior
distribution $p(\theta)$
\begin{equation}
i(p(\theta)) = -\frac{d^2 \log p(\theta)}{d\theta^2}
\end{equation}
and the information of an 
$\epsilon$-information (large-variance) prior $p_0(\theta)$ with the same mean
($\overline{\theta}$) as $p(\theta)$
\[
i(p_0(\theta)) = -\frac{d^2 \log p_0(\theta)}{d\theta^2}
\]
The authors then define the \emph{ESS} as the integer $m$ that minimizes
\begin{equation}
\label{ess::MTM1}
\vert i(p_0(\overline{\theta}))+E_{Y_m} \lbrace
i_F(Y_m;\overline{\theta})\rbrace - i(p(\overline{\theta})) \vert
\end{equation}
Here, the expectation of $Y_m$ is taken over the prior-predictive
distribution under $p(\theta)$. (\ref{ess::MTM1}) is the distance
(evaluated at the prior mean $\overline{\theta}$) between the expected
posterior information for a sample of size $m$ based on the
same-mean-large-variance prior $p_0(\theta)$ (the first two terms) and the
information of the actual prior (third term).

The approach is noteworthy because it appears to be the first formal,
metric-based approach to \emph{ESS} that complies with the standard
one-parameter exponential family \emph{ESS}. Some points deserve
attention:
\begin{enumerate}
\item[(i)] Evaluating the distance (\ref{ess::MTM1}) at the mode may
  appear more natural. However, as the authors point out, only with
  the mean one obtains the one-parameter exponential family
  \emph{ESS}.
\item[(ii)] The choice of the ``same-mean-large-variance prior''
  $p_0(\theta)$ is not unique.  Yet, since the prior $p_0(\theta)$
  carries very little information, one would expect the consequences
  to be minor. For example, for Poisson data with hazard $\theta$,
  conjugate $\mbox{Gamma}(a,b)$ prior, and $p_0(\theta)$ chosen as
  $\mbox{log-normal}(m_0,s_0^2)$, the following holds:
  $\overline{\theta}=a/b=\exp(m_0+s_0^2/2)$, and for
  $m_0=\log(\overline{\theta})-s_0^2/2$ and increasing $s_0$,
  $i(p_0(\overline{\theta})) \rightarrow -1.5/\overline{\theta}^2$. This
  implies
  $ESS_{MTM} = (a-1)/\overline{\theta}+1.5/\overline{\theta} = b(1+0.5/a)$.
  The increase from the standard $ESS=b$ will be small except for
  very small $a$.
\item[(iii)] Restricting $m$ to integers seems not important; one may
  minimize the distance (\ref{ess::MTM1}) over continuous $m$ and then round
  to integers. Setting (\ref{ess::MTM1}) to zero and noting that
  $E_{Y_m}\lbrace i_F(Y_m;\overline{\theta})\rbrace = m \cdot
  E_{Y_1}\lbrace i_F(Y_1;\overline{\theta}) \rbrace$, it follows that
\begin{equation}
\label{ess::MTM2}
ESS_{MTM} =
\frac{i(p(\overline{\theta}))-i(p_0(\overline{\theta}))}{E_{Y_1}\lbrace i_F(Y_1;\overline{\theta})\rbrace}
\end{equation}
Moreover, since $p_0(\theta)$ is not unique and really only needed to
nudge the computation of the expected Bayesian posterior information
with an ``uninformative prior'', a simplified version that ignores
this prior could be used.  Additionally approximating the expected
posterior information by $m \cdot i_F(\theta)$ 
and using the prior mode $\widetilde{\theta}$ instead of the prior mean
$\overline{\theta}$ leads to the \emph{ESS}
suggested by Pennello and Thompson (2008) 
\begin{equation}
\label{ess::MTMP}
ESS_{MTM.P} = \frac{i(p(\widetilde{\theta}))}{i_F(\widetilde{\theta})}
\end{equation}
which will usually be easier to compute than (\ref{ess::MTM2}).
\end{enumerate}
Finally, (\ref{ess::MTM2}) and (\ref{ess::MTMP}) appear similar to the
precision-ratio \emph{ESS} (\ref{ess:vr::pr}). That these similarities
can be illusory will be shown in Section \ref{sec::examples}.

\subsection{The expected local-information-ratio (ELIR) ESS}
\label{sec::elir}
We propose yet another information-based \emph{ESS}, which will be shown to
be superior to current versions. The \emph{expected
  local-information-ratio (ELIR)} method also uses the prior and Fisher
information. However, instead of locally evaluating the respective information
ratio at the mean (or mode), it is defined as the mean of the prior
information to Fisher
information ratio $r(\theta)$
\begin{equation}
\label{ess::ELIR}
ESS_{ELIR} = E_{\theta} \lbrace r(\theta) \rbrace
= E_{\theta} \left\{ \frac{i(p(\theta))}{i_F(\theta)} \right\}
\end{equation}
First, and importantly, $ESS_{ELIR}$ gives the well-known effective
sample sizes for some standard one-parameter exponential families. In
Table \ref{tab::ELIR::expfamily}, the main quantities and $ESS_{ELIR}$
are shown for the mean parameter as well as the natural parameter.

\begin{table}
\small\sf\centering
\caption{Prior information $i(p(\mu))$ and $i(p(\eta))$, Fisher unit
  information $i_F(\mu)$ and $i_F(\eta)$, local-information
  ratio $r(\mu)$ and $r(\eta)$, and expected local-information-ratio $ESS_{ELIR}$ for some
  one-parameter exponential families: $\mu$ and $\eta$ are the mean and natural parameter, respectively.}
\begin{tabular}{ccccc}
parameter  & 
$i(p(\mu)), i(p(\eta))$ & 
$i_F(\mu), i_F(\eta)$ & 
$r(\mu), r(\eta)$ &  
$ESS_{ELIR}$
\\[1mm]
\multicolumn{5}{c}
{Normal ($\sigma$ known): 
$ \quad Y \sim N(\mu,\sigma^2), \quad 
\mu \sim N(m_0,s_0^2)$}
\\[1mm]
$\mu$ &
$s_0^{-2}$ & 
$\sigma^{-2}$ & 
$\sigma^2/s_0^2$ &
$\sigma^2/s_0^2$
\\[3mm]
\multicolumn{5}{c}
{$\mbox{Binomial:} \quad Y \sim \mbox{Bin}(\mu,1), 
\quad \mu \sim \mbox{Beta}(a,b), \quad  \eta = \mbox{logit}(\mu)$}
\\[1mm]
$\mu$ &
$(a-1)/\mu^2+(b-1)/(1-\mu)^2$ &
$1/\mu+1/(1-\mu)$ &
$(a-1)(1-\mu)/\mu+(b-1)\mu/(1-\mu)$ &
$a+b \, (a,b>1)$
\\
$\eta$ & 
$(a+b) \exp(\eta)/(1+\exp(\eta))^2$ &
$      \exp(\eta)/(1+\exp(\eta))^2$ &
$a+b$ &
$a+b$
\\[3mm]
\multicolumn{5}{c}
  {$\mbox{Poisson:} \quad Y \sim \mbox{Pois}(\mu), 
\quad \mu \sim \mbox{Ga}(a,b), \quad \eta = \log(\mu)$}
\\[1mm]
$\mu$ &
$(a-1)/\mu^2$ &
$1/\mu$ &
$(a-1)/\mu$ &
$b \, (a>1)$ \\
$\eta$ &
$b \exp(\eta)$ &
$\exp(\eta)$ &
$b$ &
$b$ \\[3mm]
\multicolumn{5}{c}
{$\mbox{Exponential:} \quad Y \sim \mbox{Exp}(1/\mu), 
\quad \mu \sim \mbox{Inv-Ga}(a,b), \quad \eta = 1/\mu$}
\\[1mm]
$\mu$ &
$-(a+1)/\mu^2+2b/\mu^3$ &
$1/\mu^2$ &
$-(a+1)+2b/\mu$ &
$a-1$
\\
$\eta$ &
$(a-1)/\eta^2$ &
$1/\eta^2$ &
$a-1$ &
$a-1$ \\[3mm]
\multicolumn{5}{c}
{$\mbox{Chi-square:} \quad s_d^2 \sim 
\mbox{Ga}(d/2,d/(2\sigma^2))  
\quad \sigma^2\sim \mbox{Inv-Ga}(a,b), \quad \eta = 1/\sigma^2$}
\\[1mm]
$\sigma^2$ &
$-(a+1)/\sigma^4+2b/\sigma^6$ &
$d/(2\sigma^4)$ &
$-2(a+1)/d+4b/(2\sigma^2)$ &
$2(a-1)/d$ 
\\
$\eta$ &
$(a-1)/\eta^2$ &
$d/(2\eta^2)$ &
$2(a-1)/d$ &
$2(a-1)/d$ 
\\
\end{tabular}
\label{tab::ELIR::expfamily}
\end{table}

For the natural parameter $\eta$, $ESS_{ELIR}$ is the standard
\emph{ESS} without any boundary restriction on the parameters. Here,
the information ratio $r(\eta) = i(p(\eta))/i_F(\eta)$ does not
depend on the parameter. For the natural parameter, the
sampling and prior distribution can be written as
\[
f(y \vert \eta) = \exp \lbrace y\eta - M(\eta) \rbrace, \quad
p(\eta) = \exp \lbrace n_0m_0\eta - n_0 M(\eta) \rbrace
\]
Since $i_F(\eta) = d^2M(\eta)/d\eta^2$, it follows that
$ESS_{ELIR}=n_0$. For example, for binary data with a $\mbox{Beta}(a,b)$ prior
for the mean $\mu$, $\eta = \log \lbrace \mu/(1-\mu)\rbrace$,
$M(\eta) = \log \lbrace 1+\exp(\eta) \rbrace$, and $n_0=a+b$.
For Poisson data with a Gamma prior for the mean $\mu$,
$\eta=\log(\mu)$, $M(\eta) = \exp(\eta)$, and $n_0=b$.

While the standard effective sample sizes are obtained for the natural
parameter, some special cases arise for vague priors of the mean
parameter $\mu$ (Table \ref{tab::ELIR::expfamily}). For example, for binary
data with a Beta(a,b) prior, $ESS=a+b$ is only obtained for
$a>1,b>1$. If one of the parameters is less than 1, $ESS_{ELIR}$ is
not defined because the expectation of the local information ratio
$r(\mu) = (a-1)(1-\mu)/\mu+(b-1)\mu/(1-\mu)$ does not exist; for the
uniform distribution $(a=b=1)$, $ESS_{ELIR}=0$; and, finally $a=1,
b>1$ (or $a>1, b=1$) leads somewhat suprisingly to $ESS_{ELIR}=1$,
since (for the former) $ESS_{ELIR} = (b-1) E_{\mu} \lbrace \mu/(1-\mu)
\rbrace = (b-1)a/(b-1) = a = 1$.


\subsection{Examples}
\label{sec::examples}
We now discuss two examples with non-conjugate prior distributions and
show that the \emph{ESS} for the methods discussed so far can differ
considerably.
\subsubsection{Normal data with a Student-t prior}
We first assume normal data (with known variance $s^2$) and a
Student-t prior with $df$ degrees of freedom for the mean parameter
$\theta$.  Such a heavy-tailed prior is robust in the sense that the
prior influence decreases with increasing conflict between the data
and the prior (O'Hagan (1979), O'Hagan and Pericchi (2012)). The
Fisher and prior information are
\[
i_F(\theta) = 1/s^2, \quad 
i(p(\theta)) = \frac{df+1}{df}
\frac{1-\theta^2/df}{(1+\theta^2/df)^2}
\]
Noting that the prior information for a t-prior with scale $s_0$ is
$i(p(\theta))/s_0^2$, the variance-ratio and precision-ratio \emph{ESS} are 
\[
ESS_{VR}=ESS_{PR}= (s/s_0)^2(df-2)/df \qquad (df>2)
\]
On the other hand, using a large-variance (in the limit improper)
prior $p_0(\theta)$,
\[
ESS_{MTM}= (s/s_0)^2(df+1)/df \qquad (df>1)
\]
and $ESS_{MTM.P}=ESS_{MTM}$.
Finally, intergrating $i(p(\theta))$ over the prior distribution gives
\[
ESS_{ELIR}=(s/s_0)^2(df+1)/(df+3)
\]
Table \ref{tab::ess::examples} (upper part) shows that for small $df$,
the interesting case if robustness is the aim, these \emph{ESS} differ
considerably. This is problematic when deciding on the size of an
experiment that will incorporate prior information in the analysis. Of
note, the above formulas show that with increasing degrees of freedom,
\emph{ESS} is increasing when using the inverse of the variance
($ESS_{VR}, ESS_{PR}$) but decreasing when using the curvature at the
mean ($ESS_{MTM}$). For $ESS_{ELIR}$, which uses the expected
curvature, \emph{ESS} is increasing.

\begin{table}
\small\sf\centering
\caption{Prior ESS for various methods: for normal
  data (known $s^2=100$) with Student-t(df) prior, and for
  exponential data with generalized Gamma(a,s=1,f) prior.}
\begin{tabular}{crrrrrrr}
         &    & & & & & & \\
 \multicolumn{8}{c}{ESS for normal data with Student-t prior} \\[1mm]
         & \multicolumn{2}{c}{df} & VR   & PR     & MTM & MTM.P & ELIR  \\
         & \multicolumn{2}{c}{2}  & ---  & ---    & 150    & 150      & 60 \\
         & \multicolumn{2}{c}{3}  & 33   & 33     & 133    & 133      & 67 \\
         & \multicolumn{2}{c}{4}  & 50   & 50     & 125    & 125      & 71 \\
         & \multicolumn{2}{c}{5}  & 60   & 60     & 120    & 120      & 75 \\
         & \multicolumn{2}{c}{10} & 80   & 80     & 110    & 110      & 85 \\
         & \multicolumn{2}{c}{50} & 96   & 96     & 102    & 102      & 96 \\[2mm]
 \multicolumn{8}{c}{ESS for exponential data with generalized-Gamma
   prior (a,s=1,f)} \\[1mm]
distribution & $a$   & $f$ & VR   & PR     & MTM & MTM.P & ELIR  \\
Gamma    &  9.00   &   1.00 &  10.0 &   6.2 &  9.0 &  8.0 &  8.0  \\
Weibull &  3.00  &   3.00 &   8.6 &   3.5 &  7.3  &  6.0 &  8.0 \\ 
gen-Gamma &  2.54  &   3.54 &   7.9 &   2.3 &  6.4  &  5.4 &  8.0 \\[2mm]
Gamma    &  25.00 &  1.00 &  26 &  22 &  25 &  24 &  24 \\ 
Weibull  &  5.00  &   5.00&   20 &   15 &  18 &  20 &  24 \\
gen-Gamma &  4.52  &  5.52 &  19  & 14&   16 &  19 &  24  \\[2mm]
Gamma    &  49.00 &   1.00 &  50 &  46 &  49 &  48 &  48 \\
Weibull  &   7.00 &   7.00 &  36 &  32 &  33 &  42 &  48 \\
gen-Gamma &   6.52  &   7.52 &  35 &  30 &  31 &  41 &  48 \\[2mm]
Gamma    &   81.00  &   1.00 &  82 &  78 &  81 &  80 &  80 \\
Weibull  &    9.00  &   9.00 &  58 &  53 &  53 &  72 &  80 \\ 
gen-Gamma &    8.51  &  9.51  & 55  & 51  & 50  & 71  & 80  \\[2mm]
Gamma    &  121.00  &  1.00 & 122 & 118 & 121 & 120 & 120 \\
Weibull  &   11.00  & 11.00 &  84 &  79 &  77 &  110 & 120 \\
gen-Gamma &   10.51  & 11.51 &  81 &  76 &  74 &  109 & 120 \\[2mm]
Gamma    &  169.00  &   1.00 & 170 & 166 & 169 & 168 & 168 \\
Weibull  &   13.00  &  13.00 & 115 & 110 & 106 & 156 & 168 \\ 
gen-Gamma &   12.51  &  13.51 & 111 & 107 & 102 & 155 & 168 
\end{tabular}
\label{tab::ess::examples}
\end{table}

\subsubsection{Exponential data with a generalized Gamma prior}
The second example assumes exponentially distributed data, for which
the prior of the hazard parameter $\theta$ is a generalized Gamma
distribution with shape parameter $a$, scale parameter $s$, and
family parameter $f$. Its density is
\[
p(\theta) = \frac{f \theta^{a-1} \exp\lbrace -(\theta/s)^f) \rbrace}{s^a \Gamma(a/f)}
\]
This three-parameter distribution offers more flexibility than the conjugate Gamma
distribution and may thus be useful when representing prior
information (for example three quartiles elicited from experts). It
includes Gamma $(f=1)$ and Weibull $(f=a)$ distributions as special cases.
The Fisher and prior information are
\[
i_F(\theta) = 1/\theta^2, \quad i(p(\theta)) = (a-1)/\theta^2+f(f-1)\theta^{f-2}/s^f
\]
The variance-ratio and precision ratio \emph{ESS} are
\[
ESS_{VR} = E(\theta^2)/\mbox{Var}(\theta), \quad ESS_{PR} = E^{-1}(1/\theta^2)/\mbox{Var}(\theta)
\]
which follow from $E(\theta^r) = s^r\Gamma \lbrace (a+r)/f \rbrace /\Gamma(a/f)$;
note that $ESS_{PR}$ only exists for $a>2$.
Further, using a large-variance Gamma prior for $p_0(\theta)$,
\[
ESS_{MTM} = a + f(f-1)\left[\Gamma \lbrace (a+1)/f \rbrace \right]^f /
\lbrace \Gamma(a/f) \rbrace ^f
\]
Using the mode $\widetilde{\theta} = s \lbrace (a-1)/f \rbrace^{1/f}$,
the simplified $ESS_{MTM.P}$ is 
\[
ESS_{MTM.P} = af-f
\]
Finally, the expected local-information-ratio \emph{ESS} is 
\[
ESS_{ELIR} = af-1
\]
Table \ref{tab::ess::examples} (lower part) shows the effective sample
sizes for some parameter constellations (including Gamma, Weibull, and
genuine generalized Gamma distributions), which have been grouped by
equal $ESS_{ELIR}$ values. The dilemma is the same as in the first
example: the \emph{ESS} can differ considerably, in particular with
increasing amounts of prior information.

\subsection{The predictive consistency criterion}
So far we have discussed various approaches to \emph{ESS}, which work
well for conjugate settings but can differ considerably otherwise. To
resolve  this dilemma, more than fulfilling the exponential family
criterion is needed. We require the \emph{ESS} to meet the additional
\emph{predictive consistency criterion}: 
\begin{center}
  Predictive consistency: for a sample of size $N$, the expected \\
  posterior \emph{ESS} must be the sum of the prior \emph{ESS} and
  $N$.
\end{center}

For some of the examples of Section \ref{sec::examples}, Table
\ref{forward:ess} shows the difference between the expected posterior
\emph{ESS} and $N$, which should be the prior \emph{ESS}. Only
$ESS_{ELIR}$ seems to be predictively consistent.  It should be noted
that the results are not exact: they represent the mean of 10000
simulations, each generating data of size $N$ (10,100,1000) from the
prior predictive distribution and then obtaining the respective
posterior distribution and its \emph{ESS}.  Interestingly, for large
$N$ it seems that the difference of the expected posterior \emph{ESS}
and $N$ converges to $ESS_{ELIR}$ for all methods.
 
\begin{table}[h]
\small\sf\centering
\caption{Prior ESS and expected posterior $ESS-N$ for planned
  sample sizes N = 10, 100, and 1000: for normal data
  (variance=100) with Student-t prior  and exponential data with
  Weibull prior (see Section \ref{sec::examples})} 
\begin{tabular}{cccccc}
   \multicolumn{6}{c}{} \\
   &  \multicolumn{5}{c}{$Y \vert \theta \sim N(\theta,10^2), \quad \theta \sim \mbox{Student-t}(df)$ } \\[1mm]
   &  & prior ESS &  \multicolumn{3}{c}{(expected posterior ESS)--N} \\
   & method &  &  N=10 & N=100 &  N=1000 \\[1mm]
df=2 & VR & --- &  36  & 54   & 59 \\
     & MTM & 150 &  88 &  78    & 62 \\
     & MTM.P & 150 &  137 &  85    & 62 \\
     & ELIR &  \bf{60} &  \bf{60} &  \bf{60} &  \bf{60} \\[1mm]
df=3 & VR & 33  & 48   &  62  & 67 \\
     & MTM & 133 &   110 &  83    & 70 \\
     & MTM.P & 133 &   125 &  87    & 70 \\
     & ELIR & \bf{67}  &  \bf{67} &  \bf{67} &  \bf{68} \\[1mm]
df=4 & VR & 50 & 57   &  68  & 70 \\
     & MTM & 125 &   112 &   88    & 73 \\
     & MTM.P & 125 &   119 &   90    & 73 \\
     & ELIR &  \bf{71} &  \bf{72} &  \bf{72} &  \bf{71} \\[1mm]
df=5 & VR & 60 & 63   & 72   & 75 \\
     & MTM & 120 &  112 &  89    & 77 \\
     & MTM.P & 120 &  115 &  91    & 77 \\
     & ELIR &  \bf{75} &  \bf{75} &  \bf{75} &  \bf{75} \\[1mm]
df=10 & VR & 80 &  80  & 83   & 85 \\
     & MTM & 110      &  107 &   94    & 86  \\
     & MTM.P & 110      &  107 &   95    & 86  \\
     & ELIR & \bf{85}  &  \bf{85} &  \bf{85} &  \bf{85} \\[1mm]
df=50 & VR  &   96      &    96    &  96       & 96 \\
     & MTM  &  102      &   101    &  99       & 97  \\
     & MTM.P  &  102      &   99    &  99       & 97  \\
     & ELIR & \bf{96}   &  \bf{96} &  \bf{96}  &  \bf{96} \\[3mm]
     & \multicolumn{5}{c}{$Y \vert \theta \sim \mbox{Exp}(\theta),
     \quad \theta \sim \mbox{Weibull}(a,s)$} \\[1mm]
   &  & prior ESS &  \multicolumn{3}{c}{(expected posterior ESS)--N} \\
      & method &  &  N=10 & N=100 &  N=1000 \\[1mm]
a=3 & VR    &  8.6 & 9.6 &   10 &  10 \\
& PR    &  3.6 &  5.5 &  6.0   & 6.2 \\
& MTM   &  7.3 &  8.2 &  8.8   & 9.9 \\
& MTM.P &  6 &    6.8 &  7.7 &   7.9 \\
& ELIR  &  \bf{8} &    \bf{8.0} &  \bf{7.9} &  \bf{8.0} \\[1mm]
a=5 & VR    & 20 &   23 & 26 &  26\\
& PR    & 15 &   19 & 22 &  22\\
& MTM   & 18 &   20 & 24 &  24\\
& MTM.P & 20 &   20 & 22 &  23\\
& ELIR  & \bf{24} &   \bf{24} & \bf{24} &  \bf{24}\\[1mm]
a=7 & VR    & 36 &   41 & 49 &  50\\
& PR    & 32 &   37 & 45 &  46\\
& MTM   & 33 &   37 & 44 &  48\\
& MTM.P & 42 &   41 & 44 &  47\\
& ELIR  & \bf{48} &   \bf{48} & \bf{48} &  \bf{48}\\[1mm]
a=9 & VR    & 58 &   64 & 77 &  84\\
& PR    & 53 &   59 & 73 &  80\\
& MTM   & 53 &   57 & 69 &  79\\
& MTM.P & 72 &   69 & 71 &  78\\
& ELIR  & \bf{80} &   \bf{80} & \bf{80} &  \bf{81}\\[1mm]
a=11 & VR    & 84 &   91 & 111 & 123\\
& PR    & 79 &   86 & 107 & 119\\
& MTM   & 77 &   82 & 99  & 117\\
& MTM.P & 110 &  107 & 105 & 116\\
& ELIR  & \bf{120} &  \bf{120} & \bf{120} & \bf{121}\\[1mm]
a=13 & VR    & 115 &  123 & 148 & 168\\
& PR    & 110 &  118 & 144 & 164\\
& MTM   & 106 &  111 & 133 & 158\\
& MTM.P & 156 &  152 & 146 & 158\\
& ELIR  & \bf{168} &  \bf{168} & \bf{167} & \bf{166}
\end{tabular}
\label{forward:ess}
\end{table}

In fact, it can be shown that $ESS_{ELIR}$ is predictively consistent
for any planned sample size $N$. The proof is as follows: let $Y_N$
be the predictive data of size $N$ with posterior distribution
$p(\theta \vert Y_N)$, for which the posterior $ESS_{ELIR}$ is
\[
E_{\theta \vert Y_N} \left\{
\frac{i(p(\theta))-d^2\log f(Y_N \vert \theta)/d\theta^2}
{i_F(\theta)}
\right\}
\]
The expected posterior effective sample size under the prior
predictive distribution is then 
\[
E_{Y_N} \left[
E_{\theta \vert Y_N}  
\left\{
\frac{i(p(\theta))-d^2\log f(Y_N \vert \theta)/d\theta^2}
{i_F(\theta)}
\right\}
\right]
\]

\[
= E_{\theta}  
\left[
  E_{Y_N \vert \theta} \left\{
\frac{i(p(\theta))-d^2\log f(Y_N \vert \theta)/d\theta^2}
     {i_F(\theta)}
\right\}
    \right]
\]

\[
= E_{\theta}  
\left\{
     \frac{i(p(\theta))+N i_F(\theta)}
     {i_F(\theta)}
\right\}
 = ESS_{ELIR}+N
\]

\subsection{Computations}

Computing $ESS_{ELIR}$ (\ref{ess::ELIR}) of a prior analytically was
possible in the examples of Section \ref{sec::examples}. For priors
derived from historical data, obtaining $ESS_{ELIR}$ analytically will
usually not be possible, except for special cases with known (or
assumed) variance components in the hierarchical model or a known
power parameter for power priors (Pocock (1976), Chen and Ibrahim
(2000), Neuenschwander, Branson, and Spiegelhalter (2009)).

If the prior \emph{ESS} cannot be computed analytically, approximations can
be used. First, if the prior is parametric, the information
$i(p(\theta))$ may be available analytically but the expectation
(\ref{ess::ELIR}) may require numerical integration or simulations
from $p(\theta)$ to obtain the empirical mean as an estimate of
$ESS_{ELIR}$.

A second approximation will be needed if $p(\theta)$ is not directly
available. For example, $p(\theta)$ may be a large simulation sample
(typically from an MCMC analysis); see Section \ref{sec::applications}
for two such applications.  While inconvenient, this does not pose serious
problems because $p(\theta)$ can be approximated by a mixture of
standard distributions (e.g., normal, Beta, Gamma) to any degree of
accuracy (Dallal and Hall (1983), Diaconis and Ylvisaker (1984)), and the
respective information $i(p(\theta))$ follows from the second
derivatives of the log-mixture distribution (see Appendix).

In this context, it should be noted that the \emph{ESS} of a mixture
distribution is not the respective weighted average of the
component-wise \emph{ESS}, not even for the conjugate cases of Section
\ref{sec::conjugacy}. For example, for normal data with known variance
$s^2=100$ and a mixture prior for the mean $\theta$,
$p(\theta) = 0.5 \times N(-2,2^2) + 0.5 \times N(2,2^2)$, the weighted
\emph{ESS} is $0.5\times 100/4+0.5\times 100/4=25$. The other methods
give $ESS_{VR}=ESS_{PR}=100/\mbox{Var}(\theta)=100/8=12.5$, $ESS_{MTM}=0$
(because the prior curvature at the mean is 0 for this special
mixture distribution), whereas the predictively consistent
\emph{ESS} is $ESS_{ELIR}=13.7$.

\section{Applications}
\label{sec::applications}
\subsection{Prior ESS for a proof-of-concept trial using historical control data}
\label{sec::application1}
We now discuss a recent randomized proof-of-concept phase
II trial in which the prior for the parameter in the control group was
informed by historical data.
Proof-of-concept trials aim to provide initial evidence of efficacy
for a new treatment. They increasingly use Bayesian approaches for
design and analysis (Fisch, Jones, Jones, Kerman et
al. (2015)).  Baeten, Baraliakos, Braun, Sieper et
al. (2013) describe such a trial where patients with
ankylosing spondylitis, a chronic inflammatory disease, were
randomized to the monoclonal antibody \emph{secukinumab} (n=24) or to
placebo (n=6). The Bayesian primary analysis leveraged historical
placebo data, which allowed the investigators to allocate fewer
patients to placebo. This reduced costs and trial duration and also
facilitated recruitment.

The primary efficacy endpoint was binary (response at week six).
Eight historical randomized placebo-controlled clinical trials
provided data on the placebo response rate (Table \ref{tab::AS}).  The
authors used the meta-analytic-predictive (MAP) approach
(Spiegelhalter et al. (2004), Neuenschwander et al. (2010), Schmidli,
Gsteiger, Roychoudhury, O'Hagan et al. (2014)) to quantify the
historical placebo information.

The number of responders in the
placebo group of the $j$-th historical trial is
$r_j \vert \pi_j \sim Bin(\pi_j, n_j), \quad j=1,\ldots,8$,
where $\pi_j$ is the true placebo response rate and $n_j$ the number
of patients in the placebo group. Table \ref{tab::AS} shows the fairly
heterogeneous historical placebo data, with observed response rates in the
range of 12\% (trial 7) to 37\% (trial 3).

\begin{table}[h]
\small\sf\centering
\caption{Historical ankylosing spondylitis data, summaries and mixture
  approximations of meta-analytic-predictive(MAP) prior, and $ESS_{ELIR}$}
\begin{tabular}{cccccccc}
 & & & & & & & \\
\multicolumn{8}{c}{Historical data} \\
\multicolumn{8}{l}{
23/107 (21\%), 
12/44  (27\%), 
19/51 (37\%),  
 9/39 (23\%),  
39/139 (28\%), 
6/20 (30\%),   
9/78 (12\%),   
10/35 (29\%)
} \\[2mm]
     &  &  \multicolumn{2}{c}{MAP prior}  &  &      &     &  \\
     &  &  &  & mean &   sd   &  2.5\%-97.5\%   &  \\
     &  & &     & 0.26 & 0.084  &  0.11-0.46 &   \\[2mm] 
\multicolumn{8}{c}{Approximations (single Beta and mixtures)} \\
          & $w_1 \,\, (a_1,b_1)$ & $w_2 \,\, (a_2,b_2)$ & $w_3 \,\, (a_3,b_3)$ & mean &   sd   &  2.5\%-97.5\%   & $ESS_{ELIR}$ \\
Beta      & 1.00 \, (6.8,19.7) &  &    & 0.26 & 0.083  &  0.11-0.43 & 26  \\
2-comp Beta & 0.66 \, (16.7,51.1) & 0.34 \, (3.4,9.0) &  & 0.26 & 0.084  &  0.11-0.47 & 36  \\
3-comp Beta & 0.62  \, (6.0,17.7) & 0.34 \, (36.0,110) & 0.04 \, (2.5,4.1)  & 0.26 & 0.085  &  0.11-0.45 & 38 
\end{tabular}
\label{tab::AS}
\end{table}

Denoting the placebo response rate in the new trial by $\pi_\star$ and
using the log-odds transformation, $\theta = \log \{ \pi /(1-\pi) \}$,
the simplest MAP approach assumes exchangeable parameters,
$\theta_{\star},\theta_1,\ldots,\theta_8 ~|~ \mu , \tau \sim
N(\mu,\tau^2).$ Here, the between-trial standard deviation $\tau$
characterizes between-trial heterogeneity, that is, the extent to
which the trial parameters can deviate from the mean $\mu$.

The Bayesian MAP analysis requires priors for the parameters $\mu$ and
$\tau$. For $\mu$, a vague prior is typically used. However, more care
is needed for $\tau$, in particular for few historical trials. For
example, the still popular uniform priors with large upper bound will
essentially disregard the historical data because they put too much
probability mass on unrealistically large between-trial standard
deviations. Here, we use a half-normal prior (Spiegelhalter et
al. (2004)), which puts most of its probability mass on realistic
between-trial heterogeneities ($\tau < 2$ on the log-odds scale). In
the following, we use a $N(0,10^2)$ prior for $\mu$ and a half-normal
prior with scale 1 for $\tau$. Alternatively, t-priors for $\mu$ and
half-Cauchy or half-t priors for $\tau$ could be used (Gelman (2006),
Polson and Scott (2012)).

\begin{figure}[h]
\begin{center}
\includegraphics[width=0.9\textwidth]{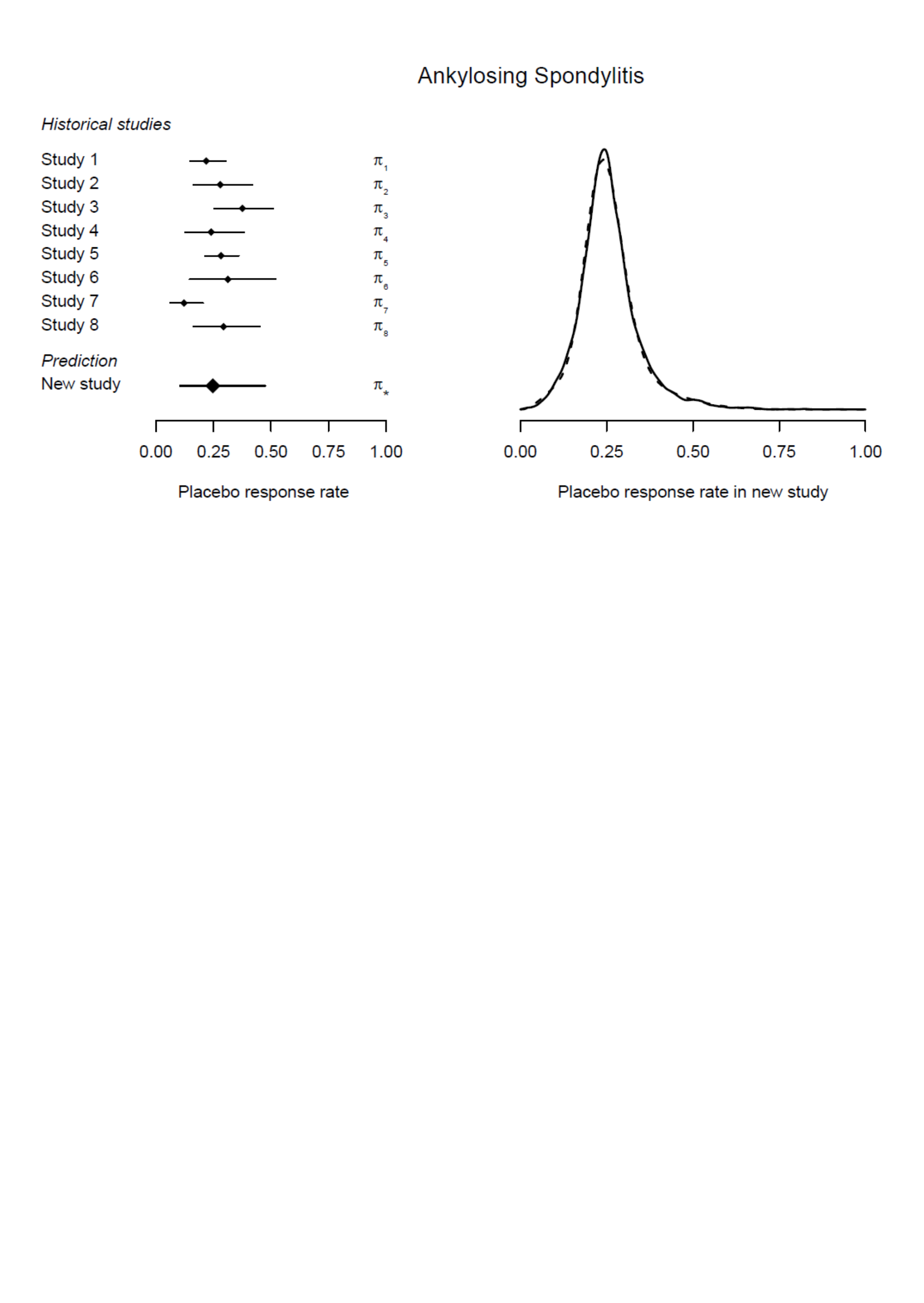}
\end{center}
\caption{
 Median and 95\%-intervals for event rates of historical 
ankylosing spondylitis trials and MAP event rate for new trial (left
panel), 
and  MAP prior density (solid line) with two-component Beta mixture approximation (dashed
line) (right panel).
}
\label{figure1}
\end{figure}

Markov chain Monte Carlo (MCMC) have been used to simulate from the
posterior distribution
$p_{MAP}(\pi_{\star}) = p(\pi_{\star} ~ \vert ~ r_1,\ldots, r_8)$,
which is summarized in Figure \ref{figure1} and Table \ref{tab::AS}.
It should be noted that MCMC provides only a simulation sample but no
analytic solution for the MAP prior. This complicates both the
calculation of the prior \emph{ESS} and the Bayesian analysis at the
end of the trial. Both issues can be addressed by approximations to
the MAP prior.  Baeten et al. (2013) used a single Beta distribution, but
mixture distributions will usually provide much better approximations
(Schmidli et al. (2014), Weber (2019), Weber et al. (2019)). Here, a mixture of two Beta
distributions (Table \ref{tab::AS}) already provides a very good fit;
Figure \ref{figure1} displays the MAP prior (density plot) and the
two-component mixture approximation.

For the design of the new trial, which aims for a smaller placebo
group by leveraging the historical information, knowing the \emph{ESS}
is important.  For the two- and three-component mixture approximation,
which give a very similar fit, the $ESS_{ELIR}$ are 36 and 38,
considerably larger than the \emph{ESS=25} from a single Beta
approximation. Of note, the predictively inconsistent $ESS_{VR}$ is 26
for all approximations, whereas the $ESS_{MTM}$ for the single Beta
and the two mixture approximations are 26, 57, and 91, respectively.

\subsection{Posterior ESS for hierarchical subgroup analyses}
\label{sec::application2}
The aim of this section is to use effective sample sizes to quantify
the gain of information for hierarchical subgroup
analyses. Hierarchical models enable sharing information across
similar but non-overlapping subgroups, which can be particularly
helpful for small subgroups.

We use the phase II trial by Chugh, Wathen, Maki, Benjamin et
al. (2009) who assessed the effect of \emph{imatinib} in ten
histological subtypes of sarcoma. The data are shown in Table
\ref{tab::ESSsubgoups}: 179 patients were available for analysis, with
sample sizes ranging from 2 to 29 for the ten subtypes. Observed
response rates for clinical benefit response (CBR) varied from 0\% for
subtypes 2 and 9 to 24\% for subtype 5.

The trial design (Thall, Wathen, Bekele, Champlin et al. (2003)) was
based on a standard hierarchical model, which exploits the anticipated
similarity of responses for the 10 subtypes.  Robust extensions of
this standard model have been discussed by Leon-Nevelo, Bekele,
M\"{u}ller, Quintana et al. (2012) and Neuenschwander, Wandel,
Roychoudhury, Bailey (2015). In the sequel, $ESS_{ELIR}$ for each
subgroup will be given for the full exchangeability model and three
robust extensions.

For binomial data, $r_j \vert \pi_j \sim Bin(\pi_j,n_j)$, a convenient
hierarchical model assumes a normal random-effects distribution for
the subtype-specific log-odds parameters
$\theta_j = \log(\pi_j/(1-\pi_j)), j=1 \ldots 10$, i.e.,
$\theta_j \vert \mu,\tau \sim N(\mu,\tau^2)$. The following prior
distributions will be used: a vague normal distribution with mean 0
and standard deviation 2 for $\mu$, and a half-normal distribution
with scale parameter 1 for $\tau$; the 95\%-interval of the latter is
(0.03,2.24), which covers very small to very large between-subtype
heterogeneity on the log-odds scale. In addition, we use
three robust mixture models, assuming the first component as
$\theta_j \vert \mu,\tau \sim N(\mu,\tau^2)$ (with priors as above)
and the second component as $\theta_j \sim N(0,2^2)$ for all
subgroups. For each subgroup, mixture weights will be 0.9/0.1,
0.75/0.25, or 0.5/0.5. The four models will be denoted by (HM-100) for
the full exchangeability model and HM-90, HM-75, HM-50 for the three
mixture models.

Here, we are interested in the \emph{ESS} relative to parallel
binomial experiments for each subtype, for which the Fisher
information is
$i_F(\theta_j) = \exp(\theta_j)/\lbrace 1+\exp(\theta_j)\rbrace ^2$.
The MCMC posterior distributions $p(\theta_j\vert r_1,\ldots,r_{10})$
have been approximated by a mixture of four Beta distributions.

For the four models, Table \ref{tab::ESSsubgoups} shows $ESS_{ELIR}$
for the response rate in each subgroup. Of course, the information
gain relative to the subgroup sample sizes $n_j$ is the largest for
the full exchangeability model (HM-100), with \emph{ESS} between 54
and 78. On the other hand, the \emph{ESS} for the robust mixture
extensions can be considerably smaller (in particular for model HM-50)
but are still much larger than the subgroup sample sizes $n_j$. 

The results show that even under robust borrowing hierarchical model
analyses for subgroups can lead to substantive information gains
compared to stratified analyses. Finally, it should be noted that for
the full exchangeability model, the posterior mixture weights increase
for all subtypes, which justifies the full exchangeability design used
in the actual trial (Thall et al. (2003)).

\begin{table}[h]
\small\sf\centering
  \caption{Sarcoma subtype data (number of responders/patients)  and 
    $ESS_{ELIR}$ for each subtype from hierarchical model analyses assuming full
    exchangeability (HM-100) or robust mixtures with 90, 75, or 50\% weight
    for exchangeability (HM-90, HM-75, HM-50).}
\begin{tabular}{lcccccc}
 &       &  & HM-100 & HM-90 &  HM-75 & HM-50 \\
   Subtype        & r/n & (\%) & \multicolumn{4}{c}{ESS} \\
1. Angiosarcoma & 2/15 &(13) &   65 &  60 &  50 &  36 \\ 
2. Ewing & 0/13 & (0) &  57 &  46 &  36 &  24 \\
3. Fibrosarcoma & 1/12 & (8) &  60 &  54 &  44 &  31 \\ 
4. Leiomyosarcoma & 6/28 & (21) & 78  &  72 &  64 &  47 \\
5. Liposarcoma & 7/29 & (24) & 76  &  68 &  59 &  43 \\
6. MFH & 3/29 & (10) & 74  &  69 &  59 &  46 \\
7. Osteosarcoma & 5/26 & (19) & 79  &  73 &  64 &  47 \\
8. MPNST &  1/5 & (20) &  57 &  48 &  39 &  23 \\
9. Rhabdomysarcoma & 0/2 & (0)  &  54 &  43 &  33 &  18 \\
10. Synovial & 2/20 & (15) &  72 &  66 &  57 &  42
\end{tabular}
\label{tab::ESSsubgoups}
\end{table}


\section{Discussion}

Modern drug and health-care development tend towards better use of the
evidence, which involves using multiple data sources via meta-analytic
methods (21st Century Cures Act (2015), European Commission (2014),
European Medicines Agency (2013, 2018), FDA
(2004, 2013, 2018)).

In this regard, the effective sample size \emph{(ESS)} of
trial-external information, which contributes to the inference in the
actual trial, is an important metric. Various methods to obtain the
ESS have been discussed recently. They are similar in that they relate
the available information (formally the prior or posterior precision)
to the Fisher information. Yet the methods can give surprisingly
different \emph{ESS}.   
We have shown that the expected local-information-ratio $ESS_{ELIR}$ addresses
the limitations of current methods. Importantly, it is predictively
consistent and thus correctly quantifies the amount of information as
an equivalent number of observations. 

Our focus has been on one-dimensional parameters. Clearly, many
applied problems involve more than one parameter, for which effective
sample sizes of individual parameters (or even parameter vectors) are
of interest; for $ESS_{MTM}$ in such settings, see Morita, Thall,
M\"uller (2012) and Thall, Herrick, Nguyen, Venier et
al. (2014). Finding predictively consistent \emph{ESS} for such cases
requires further research.

\section*{Acknowledgements}
We would like to thank the two reviewers and the associate editor for
their excellent comments, which greatly helped us to improve the
manuscript.






\section*{Supporting Information}
The on-line material is available at Biometrics website on Wiley
Online Library. It contains $ESS_{ELIR}$ functions for mixtures of
normal and Beta distributions and code to reproduce the results of
Table 3 and the two applications. Note that the R-package \emph{RBesT}
(Weber (2019), Weber et al. (2019)) is required.

\appendix 
\section{}

\subsection{Information $i(p(\theta))$ of  a mixture distribution}
If the prior (or posterior distribution) is a mixture
distribution with $K$ components, $p(\theta) = \sum_{j=1}^K w_jp_j(\theta)$, its
information is 
\begin{eqnarray*}
i(p(\theta)) & = & -d_\theta^2\log p(\theta) \\
             & = & \frac{1}{p^2(\theta)} \left[ \sum_{j=1}^K w_jp_j(\theta)
                  d_{\theta}\log p_j(\theta) \right] ^2 \\
             &   & - \frac{1}{p(\theta)} \sum_{j=1}^K w_jp_j(\theta)
                   \left[ \lbrace d_{\theta}\log p_j(\theta)\rbrace^2 
                   + d_{\theta}^2 \log p_j(\theta) \right]
\end{eqnarray*}
Here, $d_{\theta}$ and $d_{\theta}^2$ denote the first and second
derivative, respectively.

\subsection{Fitting mixture distributions}
\label{fit::mix}
Various procedures are available for fitting mixture distributions, for
example the \emph{mixfit} and \emph{automixfit} functions in the
R-package \emph{RBesT} (Weber (2019), Weber et al. (2019)), or \emph{SAS PROC
  FMM} (2014). The former has been used to fit the prior
and posterior distributions of the applications in Section \ref{sec::applications}.

\subsection{R-function for the ESS of a mixture of Normal or Beta distributions}
\begin{verbatim}
ess_normmix = function(normmix,
                       iF.theta = function(x) return(1),
                       n.sim = 1e6) {

  # normmix:  k-component mixture distribution, a list with k elements, 
              each a 3-vector (weight, mean, standard deviation)
  # iF.theta: the function returning the Fisher information
  # n.sim:    number of simulations to obtain the ESS empirically
  K = length(normmix)
  w = sapply(normmix, function(e) e[1])
  mn = sapply(normmix, function(e) e[2])
  sd = sapply(normmix, function(e) e[3])
  
  # simulation from mixture distribution
  mixcomp = sample( x= 1:K, size = n.sim, replace = TRUE, prob = w )
  theta = rnorm(n.sim, mn[mixcomp], sd[mixcomp])
  
  # 2nd derivates (Appendix 1)
  M = 0
  for (k in 1:K) { M = M + w[k] * dnorm(theta, mn[k], sd[k])}
  
  sum1 = 0; sum2 = 0
  for (k in 1:K) {
    sum1=sum1+w[k]*dnorm(theta,mn[k],sd[k])*(-(theta- mn[k])/sd[k]^2)
    sum2=sum2+w[k]*dnorm(theta, mn[k], sd[k])*((theta-mn[k])^2/sd[k]^4-1/sd[k]^2)
  }
  
  DlogM2 = (-sum1^2/M^2+sum2/M)/iF.theta(theta)
  ESS = mean(-DlogM2)
  return(ESS)
}
\end{verbatim}
\begin{verbatim}
ess_betamix = function(betamix, 
                       iF.theta = function(x) return(1/x/(1-x)), 
                       n.sim = 1e6) {

  # betamix:  k-component mixture distribution, a list with k elements, 
              each a 3-vector (weight, a, b)
  # iF.theta: the function returning the Fisher information
  # n.sim:    number of simulations to obtain the ESS empirically
 
  K = length(betamix)
  w = sapply(betamix, function(e) e[1])
  a = sapply(betamix, function(e) e[2])
  b = sapply(betamix, function(e) e[3])

  #simulation from mixture distribution
  mixcomp = sample( x = 1:K, size = n.sim, replace = TRUE, prob = w )
  theta = rbeta(n.sim, a[mixcomp], b[mixcomp])

  # 2nd derivates (Appendix 1)
  M = 0
  for (k in 1:K) { M = M + w[k] * dbeta(theta,a[k], b[k]) }

  sum1 = 0;   sum2 = 0
  for (k in 1:K) {
    sum1 = sum1+w[k]*dbeta(theta, a[k], b[k])*
           ((a[k] - 1)/theta-(b[k]-1)/(1 - theta))
    sum2 = sum2+w[k]*dbeta(theta, a[k], b[k])*
           (((a[k]-1)/theta-(b[k]-1)/(1-theta))^2-
           ((a[k]-1)/theta^2+(b[k]-1)/(1-theta)^2))
  }

  DlogM2 = (1/M^2*sum1^2-1/M*sum2)/iF.theta(theta)
  ESS = mean(DlogM2)
  return( ESS )
}
\end{verbatim}






\end{document}